\documentclass[usenatbib,twocolumn,useAMS]{mn2e}
\usepackage{latexsym}
\usepackage{dcolumn}
\usepackage{bm}
\input{epsf}
\newcommand{\be}{\begin{equation}}
\newcommand{\ee}{\end{equation}}
\newcommand{\ba}{\begin{eqnarray}}
\newcommand{\ea}{\end{eqnarray}}
\newcommand{\no}{\nonumber}
\newcommand{\bi}{\begin{itemize}}
\newcommand{\ei}{\end{itemize}}
\newcommand{\bfi}{\begin{figure}
\epsfxsize=9cm
\epsffile}
\newcommand{\efi}{\end{figure}}
\newcommand{\mpch}{$h^{-1}$ Mpc}
\newcommand{\muk}{\mu {\rm K}}
\newcommand{\mnras}{MNRAS}
\newcommand{\apj}{ApJ}
\newcommand{\apjl}{ApJ}

\newcommand{\prd}{PRD}

\title[Dark flow kSZ]{The dark flow induced small
  scale kinetic Sunyaev Zel'dovich effect}
\author[Zhang]{Pengjie Zhang$^1$
\\$^1$Key Laboratory for Research in Galaxies and Cosmology, Shanghai
  Astronomical Observatory, Nandan Road 80, Shanghai, 200030,
  China;\\ pjzhang@shao.ac.cn}
\begin{document}
\maketitle
\begin{abstract}
Recently \citet{Kashlinsky08,Kashlinsky10} reported a discovery of a $\sim
10^3$ km/$s$ 
bulk flow of the  universe out to $z\simeq 0.3$, through the dark flow
induced CMB dipole in directions of clusters. We point out that, if this dark
flow exists, it will also induce observable CMB temperature fluctuations at
multipole $\ell\sim 10^3$, through 
modulation of the inhomogeneous electron distribution on the uniform dark
flow. The induced  
{\it small scale} kinetic Sunyaev Zel'dovich (SZ) effect will reach $\sim
1\muk^2$ at multipole $10^3\la \ell\la 10^4$, only a factor of $\sim 2$ smaller than
the conventional   kinetic SZ effect. Furthermore,  it will be correlated with the large scale structure (LSS) and its correlation with 2MASS
galaxy distribution reaches $0.3 \mu$K at $\ell=10^3$,
under a directional dependent optimal weighting scheme. We estimate that,  
WMAP plus 2MASS should  already be able to detect this dark flow induced small
scale 
kinetic SZ effect with $\sim 6\sigma$ confidence. Deeper galaxy
surveys such as SDSS can further improve the measurement.  Planck plus
existing galaxy surveys can reach $\ga 14\sigma$ detection. Existing CMB-LSS
cross correlation measurements shall be  
reanalyzed to test the existence of  the dark flow and, if it exists,  shall
be used to eliminate possible bias on the 
integrated Sachs-Wolfe effect measurement through the CMB-LSS cross correlation. 

\end{abstract}
\begin{keywords}
(cosmology:) large-scale structure of Universe: cosmic microwave background:
  theory: observations 
\end{keywords}

\section{Introduction}
Recently, 
\citet{Kashlinsky08,Kashlinsky09,Atrio-Barandela10,Kashlinsky10} analyzed 
CMB fluctuations on directions of X-ray galaxy clusters and found a large bulk flow
with speed $v_{\rm DF}\sim 10^3$ km/$s$ to  $z_{\rm DF}\sim 0.3$, toward
direction $(l_0,b_0)\sim (290^{\circ},30^{\circ})$.  This extraordinarily  large bulk flow
(the so called dark flow)
is a severe challenge to the standard $\Lambda$CDM paradigm and has fundamental
implications on the topology of the universe, the inflationary scenario and the
nature of gravity (e.g. \citealt{Carroll08,Afshordi09,Chang09,MH09,Khoury09,
  Kashlinsky10} and references therein).  

The above dark flow measurements are under scrutiny
\citep{Keisler09,Atrio-Barandela10}. It also  requires independent
confirmations, from direct \citep{Feldman09, Watkins09} and indirect
\citep{Lavaux10} velocity measurements of nearby galaxies and nearby
supernovae \citep{Gordon08}.  In this paper, we point
out that, besides the dipole from which the dark flow is inferred, the dark
flow also induces small  angular scale  CMB temperature fluctuations. If the
amplitude of the dark flow reaches the reported value of $\sim 10^3$
km/$s$, it shall be detected by combining existing CMB and galaxy measurements.

If the matter distribution where the dark flow resides is homogeneous, the dark
flow only induces a CMB dipole.  However, from galaxy surveys, we know
that the local universe is strongly inhomogeneous in the density
distribution (e.g. \citealt{Tegmark04}).  Given these
inhomogeneities, even an uniform dark flow can induce small angular scale
temperature fluctuations in the CMB sky, through 
exactly the same mechanism of the inverse Compton scattering to generate the
conventional kinetic Sunyaev Zel'dovich (SZ) effect \citep{Sunyaev72,Sunyaev80}, 
\ba
\label{eqn:ksz}
\Theta_{\rm DF}(\hat{n})&\equiv &\frac{\Delta T_{\rm DF}(\hat{n})}{T_{\rm CMB}}=\int
n_e(\hat{n},z)\sigma_Tad\chi \frac{{\bf v}_{\rm DF}\cdot
  \hat{n}}{c}\\
&=&\left[6.1\times 10^{-6}\frac{v_{\rm DF}}{10^3 {\rm km}/s}\int_0^{z_{\rm DF}}
(1+z)^2d\tilde{\chi}\right] \cos\theta\ \no\\ 
&+& 6.1\times 10^{-6}\frac{v_{\rm DF}}{10^3 {\rm km}/s}\cos\theta\int_0^{z_{\rm DF}}
\delta_e(\hat{n},z)(1+z)^2d\tilde{\chi} \no
\ea
Here, $\cos\theta\equiv \hat{n}_{\rm DF}\cdot\hat{n}$ is the
cosine between the dark flow direction $\hat{n}_{\rm DF}$ and the direction
$\hat{n}$. $z_{\rm DF}$ is the edge of the dark flow. $n_e=\bar{n}_e(1+\delta_e)$ is the 3D free electron number density,
$\bar{n}_e$ is the mean number density and $\delta_e$ is the over-density. 
Throughout the paper we adopt the fiducial value $\Omega_bh=0.031$ and thus
neglect the prefactor $\Omega_b h/0.031$ in Eq. \ref{eqn:ksz}. $\tilde{\chi}\equiv
\chi/(c/H_0)$ is the dimensionless comoving distance in unit of the Hubble
radius $c/H_0$. Throughout the paper we  adopt a flat
$\Lambda$CDM cosmology with $\Omega_m=0.27$, 
$\Omega_{\Lambda}=1-\Omega_m$, $\Omega_b=0.044$, $\sigma_8=0.84$ and
$h=0.71$.

  The first term in the last expression of Eq. \ref{eqn:ksz} is a dipole
  term. However, the second term is not, 
despite its dipole-like prefactor $\cos\theta$.  The electron
over-density $\delta_e(\hat{n})$ has a complicated directional dependence and
is clustered at scales $\la 100$ \mpch. It is this density modulation from all
free electrons producing the small scale kinetic SZ effect, the one that we
point out and 
investigate in this paper.  It clearly differs from the CMB 
dipole induced by the Earth  motion and the CMB dipole induced by the dark flow
of uniform electron distribution, both are lacking of the $\delta_e$ modulation. 
It also  differs from the  conventional kinetic SZ effect, which is further
modulated by the non-uniform velocity \citep{Vishniac87}.  Later we will find
that the two kinetic SZ effects have different clustering behaviors for this reason. 

This dark flow induced small scale kinetic SZ effect leaves unique imprints on
the CMB sky. We will estimate its auto power spectrum and its correlation with
the large scale structure (LSS). The exact value of $v_{\rm DF}$ is highly
uncertain \citep{Kashlinsky10}. Throughout the paper, we will adopt a fiducial
value $v_{\rm DF}=10^3$ km/$s$. The results presented in this paper can be
scaled to other value of $v_{\rm DF}$ straightforwardly. Given $v_{\rm
  DF}=10^3$ km/$s$, we find that it is promising to extract the dark flow 
component through CMB-LSS correlations in an unbiased manner, if the CMB maps are properly weighted
with a directional dependent weighting factor. We estimate that existing data
may already allow for $\ga 6\sigma$ detection of this effect and provide
independent test on the dark flow  scenario.

\section{The auto power spectrum of the dark flow induced kinetic SZ effect}
The dark flow induced kinetic SZ temperature fluctuation is correlated at
small scales due to clustering of the underlying electron overdensity
$\delta_e$. We derive the resulting auto power spectrum averaged over the
survey area (the appendix \S \ref{sec:cl}), 
\ba
\label{eqn:cl}
\frac{\ell^2C_{\rm DF}(\ell)}{2\pi}&=& 3.7\times 10^{-11} \left[\frac{v_{\rm
    DF}}{10^3 {\rm km}/s}\right]^2f_{T}\no \\
&\times&\frac{\pi}{\ell}\int_0^{z_{\rm cut}}
\Delta^2_e(k=\frac{\ell}{\chi},z)(1+z)^4\tilde{\chi}d\tilde{\chi} \ .
\ea
 The expression in the integral has adopted the well known Limber
 approximation.  $\Delta^2_e(k,z)$ is the electron number density power
spectrum (variance) at scale $k$ and redshift $z$. Throughout the paper we
approximate it as $\Delta^2_e=\Delta^2_m$. Here $\Delta^2_m$ is the matter
power spectrum (variance). Since the measured matter distribution in the
nearby universe agrees with the standard $\Lambda$CDM \citep{Tegmark04}, it
allows us to adopt the CMBFAST transfer function \citep{CMBFAST}.  The nonlinear
power spectrum $\Delta^2_m$ is 
calculated by the halofit formula \citep{Smith03}. 

The suppression factor $f_T$  is
\be
\label{eqn:fT}
f_T=\left\langle \cos^2\theta\right\rangle_S\ .
\ee
The average is over all directions in the survey area.  Clearly $f_T$ is a
direction dependent quantity and varies from survey to survey. Namely, the 
dark flow induced kinetic SZ effect is statistically anisotropic, unlike other
isotropic CMB components. Eq. \ref{eqn:fT} is derived under the small angle
approximation.  Its accuracy is of the order $(\Delta
\theta)^2/4\sim (\pi/\ell)^2\ll 1$ when $\ell\geq 10$, where $\Delta \theta$
is the typical angular scale involved, $\Delta \theta\sim 2\pi/\ell$. Refer to
the appendix \S \ref{sec:cl} for more details. For a full sky survey
$f_T=1/3$. 

This characteristic directional dependence can in principle be applied to
separate the  dark flow induced kinetic SZ effect from isotropic components
(primary CMB, the thermal SZ effect, the conventional kinetic SZ effect, etc.)
in the auto correlation measurement. Alternatively one can weigh the CMB
temperature fluctuations by a directional dependent function $W(\hat{n})$. It
can be desired to amplify the dark flow component with respect to other
components.  In this case, 
\be
f_T=\langle \cos^2\theta W^2(\hat{n})\rangle_S\ .
\ee

\bfi{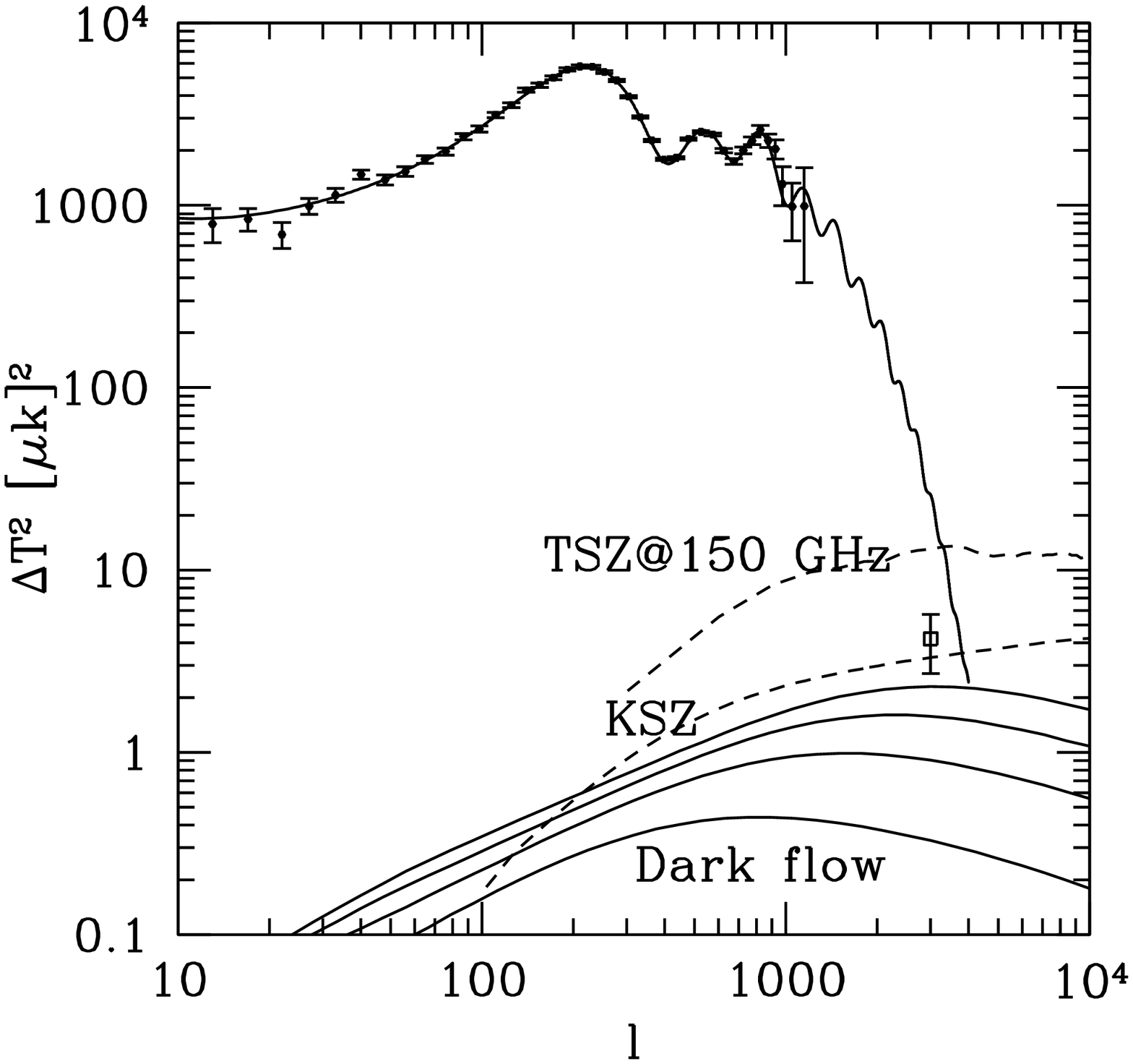}
\caption{The auto correlation power spectrum of the dark flow induced kinetic
  SZ effect. The primary CMB data points are the seven years WMAP result
  \citep{Komatsu10,Larson10}. The SZ measurement at $\ell=3000$ is from
  \citet{Lueker09}. The thermal SZ result is the one in \citet{Zhang02},
  scaled from the original  $\sigma_8=1.0$ to $\sigma_8=0.8$  with the
  $\sigma_8^7$ scaling and scaled 
  from the Raleigh-Jeans regime to $150$ GHz. The kinetic SZ result is the
  homogeneous kinetic SZ power spectrum predicted from the model of
  \citet{Zhang04}. The four solid lines are for the dark flow models with
  $z_{\rm DF}=0.1,0.2,0.3,0.4$ from bottom up.  We have adopted $v_{\rm DF}=10^3$
  km/$s$. The shown power
  spectra is for a full sky survey, with the suppression factor
  $f_T=1/3$. The SPT sky \citep{Lueker09} has $f_T\sim 0.14$, so the dark
  flow contribution to the measured SZ effect at $\ell=3000$ is  $\sim 0.3\muk^2$, sub-dominant to other components. \label{fig:ksz}}
\efi

 $C_{\rm DF}(\ell)$ shows a number of difference in the $\ell$
dependence comparing to that of the conventional kinetic SZ effect
(Fig. \ref{fig:ksz}), the major reason is that the dark flow induced kinetic
SZ fluctuation is linear in density fluctuations while the conventional one is
quadratic through the interplay between density and velocity
inhomogeneity.   Another reason is that the dark flow induced kinetic SZ
effect comes from much lower redshift than the conventional kinetic SZ
effect. Since the typical redshift of the dark flow induced kinetic SZ effect
is proportional to $z_{\rm DF}$, the peak moves to smaller scale (larger $\ell$) when
$z_{\rm DF}$ increases. For the fiducial value of $v_{\rm DF}=10^3$ km/$s$ out to
$z_{\rm DF}=0.3$,  the  
dark flow induced kSZ power spectrum, averaged over the whole sky with $W=1$,
peaks at  $\ell\simeq 2000$ with an 
amplitude $1.6 \muk^2$,  a factor of $\sim 2$ smaller than the
conventional kinetic SZ effect (e.g. \citealt{Ma02,Zhang04,HH09}) at the same
scale. Overall, this dark flow induced kinetic SZ effect is non-negligible,
ranging from $\sim 100\%$ of the conventional one at $\ell=200$ to $\sim 30\%$
at $\ell=10^4$.

The SPT collaboration  measured a combined SZ power  spectrum
(tSZ+0.46$\times$ kSZ) at $150$ GHz band and 
$\ell=3000$  to be $4.2\pm 1.5\muk^2$  \citep{Lueker09}. This
measurement is a factor of   $\sim 2$ lower than the predicted thermal SZ
effect with $\sigma_8=0.8$ (e.g. \citealt{Zhang02,Lueker09}), but
consistent with recent  WMAP and ACT measurements \citep{Komatsu10,ACT10}. 
 The dark flow induced kinetic SZ effect is $\sim 0.3\muk^2$ at the
 analyzed SZ  sky where $f_{T}\simeq 0.14$.  Given its sub-dominance to the
 conventional kinetic SZ effect and the thermal SZ effect
 (Fig. \ref{fig:ksz}),  it is difficult to perform a robust test of the reported
 dark flow against existing SZ power spectrum measurements. Future large-area SZ
 measurements at $217$ GHz,  the null of the thermal SZ effect, may be able to
 infer its existence through the directional dependence in the  $f_T$ prefactor.

\bfi{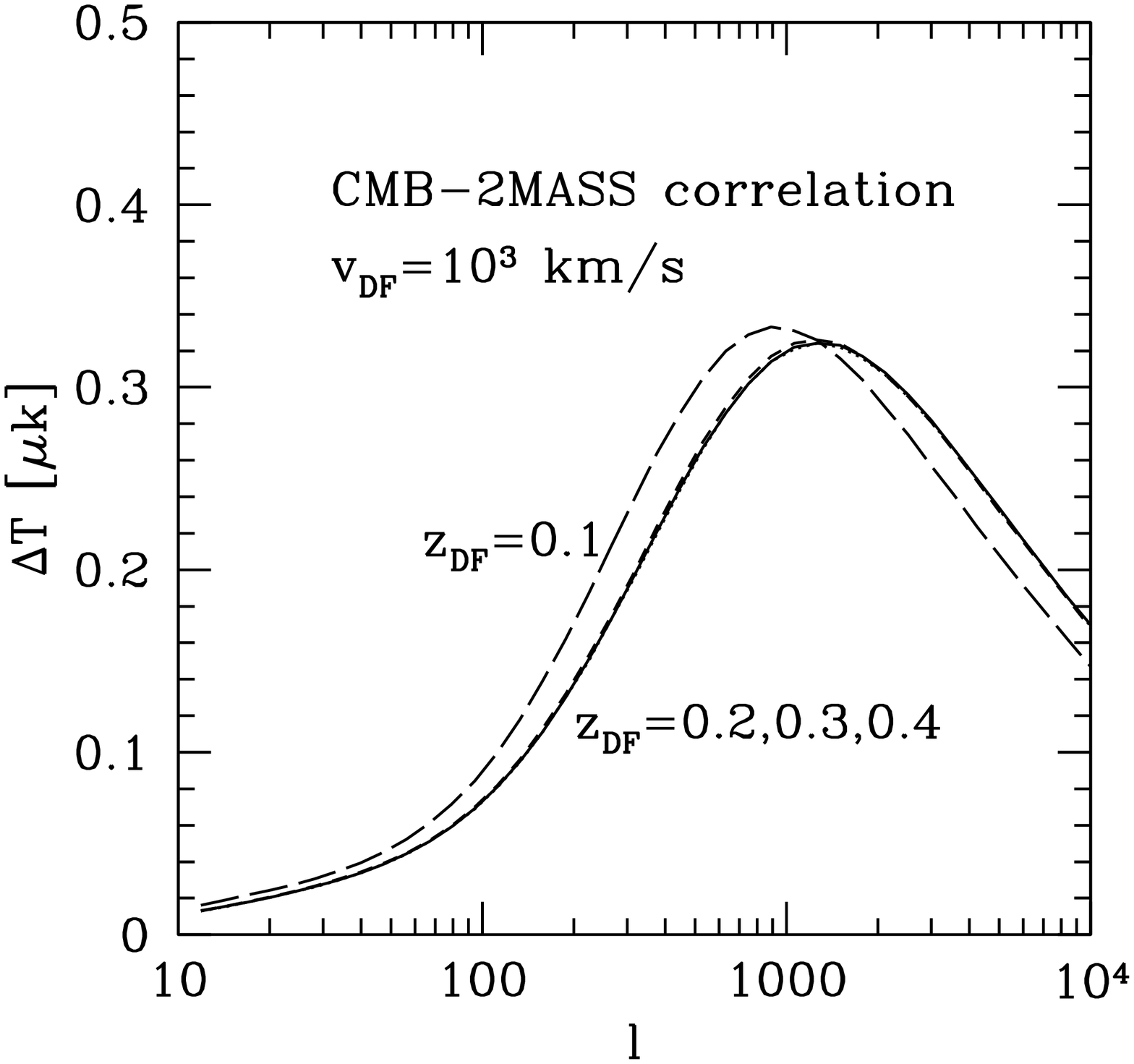}
\caption{The cross correlation power spectrum between the dark flow induced kinetic
  SZ effect and a full sky galaxy survey of 2MASS-like. A directional
  dependent weighting $W=\cos\theta$ is adopted. Since most 2MASS galaxies
  reside at $z<0.1$, the cross correlation varies little for bulk flows
  beyond $z\ga 0.1$. For this reason, the three curves of $z_{\rm
    DF}=0.2,0.3,0.4$ are barely distinguishable.  \label{fig:Tg}}
\efi

\section{Cross correlation with galaxies}
The dark flow induced kinetic SZ effect is also  correlated with tracers of the large
scale structure at $z<z_{\rm DF}$.  The resulting cross power spectrum between
the dark flow induced kinetic SZ effect and the galaxy distribution is 
\ba
\label{eqn:tg}
\frac{\ell^2C_{Tg}}{2\pi}&\simeq& 6.1\times 10^{-6} \frac{v_{\rm
    DF}}{10^3 {\rm km}/s}f_{Tg}  \\
&\times&\frac{\pi}{\ell}\int_0^{z_{\rm cut}}
\Delta^2_{eg}(k=\frac{\ell}{\chi},z)(1+z)^2\tilde{\chi}\bar{n}_g(z)dz \no\ .
\ea
$\bar{n}_g$ is the galaxy distribution function normalized that $\int
\bar{n}_g(z)dz=1$. We approximate the electron-galaxy cross power spectrum
$\Delta^2_{eg}=b_g\Delta^2_m$ where $b_g$ is the galaxy bias.  The suppression
factor  
\be
f_{Tg}=\left\langle \cos\theta W(\hat{n})\right\rangle_S\ .
\ee
Here the average is over the overlapping sky of CMB and galaxy
surveys. $f_{Tg}$ is also direction dependent and varies from survey to
survey. Especially, $f_{Tg}=0$ with $W=1$
for full sky surveys. Thus for WMAP+2MASS, we have to choose $W\neq 1$ to
avoid the cancellation. A natural choice  is $W(\hat{n})=\cos\theta$, for
which $f_{Tg}=1/3$. A nice property about this choice is that $\langle
W\rangle_S=0$ and thus isotropic components in CMB maps  (primary CMB, the
thermal  SZ, conventional kinetic SZ effect, dusty star forming galaxies, etc.) do not
bias the cross correlation.\footnote{Realistic correlation analysis masked the galactic plane, so
  $f_{Tg}\neq 0$ even for full sky surveys like WMAP+2MASS. This issue is
  important for actual data analysis. But for the theoretical study
  presented in this paper it is safe to neglect it. } 

The cross correlation signal between the dark flow induced kinetic SZ effect
and 2MASS galaxy distribution is shown in Fig. \ref{fig:Tg}.  The galaxy
distribution is adopted from \citet{Afshordi04}, along with the galaxy bias
$b_g=1.18$.  Since the galaxy distribution peaks at $z<0.1$, the peak of the
cross power spectrum is at lower $\ell$ than that in the auto power
spectrum. For a dark flow with $v_{\rm DF}=10^3$ km/$s$ out to $z_{\rm
  DF}=0.3$, the cross power spectrum peaks at $\ell\simeq 1.3\times 10^3$ with
amplitude $0.32  \muk$.  

Given the existence of this cross correlation, it is interesting to ask
whether it will impact the interpretation of existing   CMB-LSS cross
correlation measurements, whose primary goal is to detect the integrated
Sachs-Wolfe effect (ISW)  (\citet{Giannantonio08} and references therein).
Since $f_{Tg}=0$ with $W=1$  for full sky surveys,  this dark flow induced kinetic SZ 
effect is unlikely to significantly bias the integrated Sachs-Wolfe (ISW) measurement
through WMAP+2MASS  (NVSS), unless the CMB masks causes significant deviation
of $f_{Tg}\neq 0$. For surveys with partial sky coverage, $f_{Tg}\neq 0$ in
general. So the ISW measurements based on these surveys are biased by this
component.  Interestingly,  depending on the direction of survey area, this dark
flow induced kinetic SZ effect may increase or decrease the measured
correlation strength. Given the comparable correlation strength of the two
effects, this issue shall be  taken into account of future data analysis to
avoid bias in dark energy constraint, especially for low  redshift galaxy 
surveys with partial sky coverage like SDSS.

\bfi{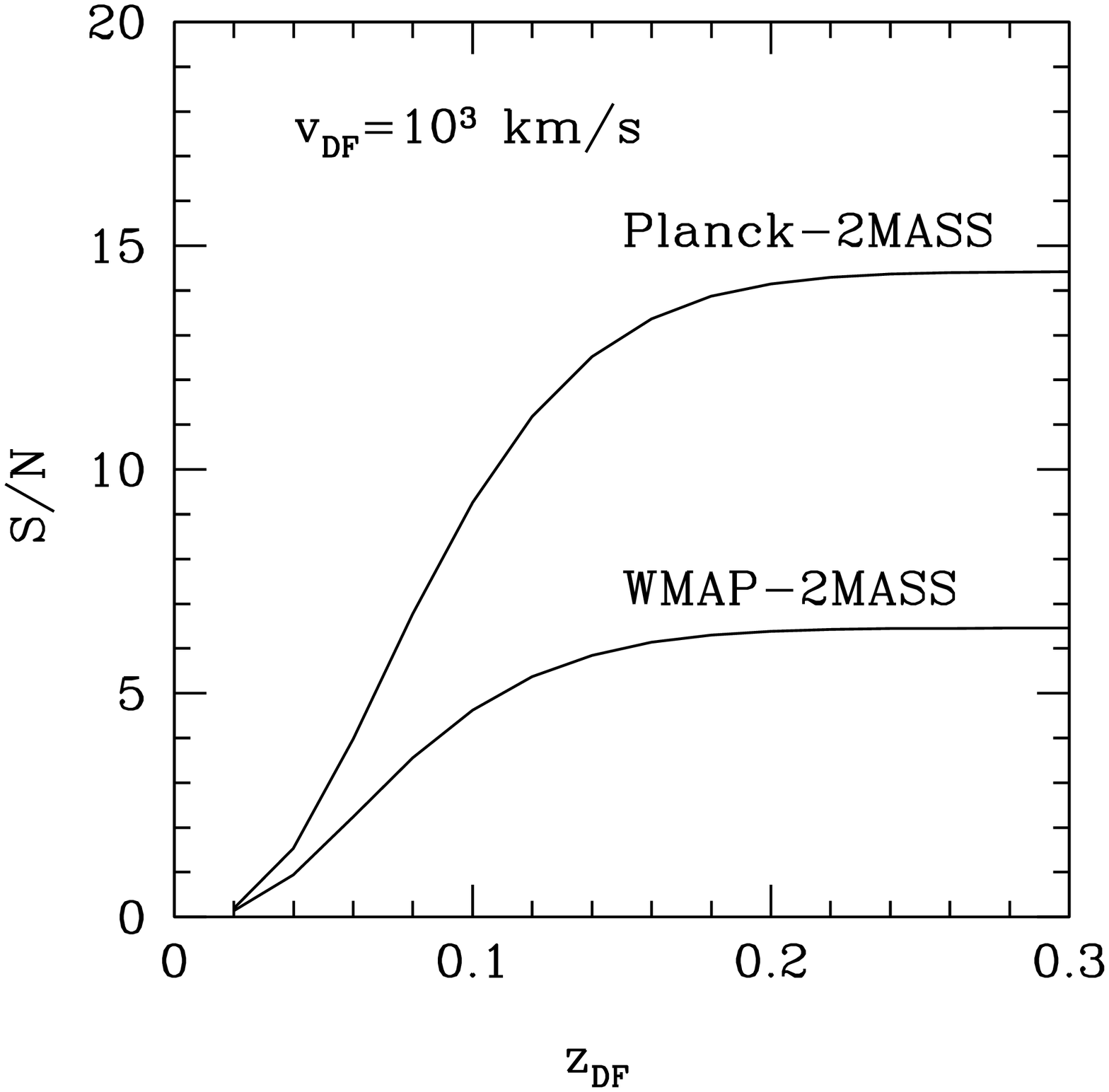}
\caption{The predicted S/N of  WMAP-2MASS (Planck+2MASS) cross correlation
  designed  for the dark flow detection, for which an unbiased optimal
  directional dependent weighing scheme is applied. WMAP+2MASS is able to
  detect the bulk flow at $\sim 6\sigma$ confidence level. Planck+2MASS is
  able to improve the S/N by a factor of $\sim 2$.  Since most 2MASS galaxies
  reside below $z=0.1$, 
  it is inefficient to detect the bulk flow beyond $z=0.1$, explaining the
  plateau in the S/N curves.   Cross correlating WMAP and Planck with deeper galaxy
  surveys such as SDSS, LAMOST and BOSS will improve the detection at
  $z>0.1$.  Thus existing surveys (WMAP, 2MASS and SDSS especially)
  are already able to detect the dark flow reported by
  \citet{Kashlinsky08,Kashlinsky10}.   Since the S/N $\propto v_{\rm DF}$, the
  cross correlation measurement of existing surveys can detect
  dark flow  with speed $\sim 500 $ km/$s$ at $\ga 3\sigma$ confidence level
  and Planck+2MASS can improve the sensitivity to $\sim 200$
  km/$s$.   \label{fig:SN}}   
\efi

Thanks to its characteristic directional dependence, at least in principle we
are able to measure the dark flow  through CMB-LSS cross correlation in a way
unbiased by other CMB components such as the ISW effect. To do so, we need to
design a weighting scheme such that  $\langle W\rangle_S=0$ to eliminate
isotropic components in CMB maps (primary CMB, the thermal 
SZ, conventional kinetic SZ effect, dusty star forming galaxies, the ISW
effect, etc.). The statistical measurement error under such weighting is
\be
\frac{\Delta C_{Tg}}{C_{Tg}}=\sqrt{\frac{1+\langle W^2\rangle_S (C^{\rm
      CMB}+C^{\rm CMB,N})(C_g+C_{g,N})/C_{Tg}^2}{2\ell\Delta \ell
    f_{\rm sky}}}
\ee
Here, $C^{\rm CMB}$ is the power spectrum of primary CMB and $C^{\rm CMB,N}$
is the one of other components including the instrumental noise.  Since
$C_{\rm DF}\ll C^{\rm CMB}+C^{\rm kSZ}+\cdots$, we can safely neglect the
contribution of $C_{\rm DF}$ in $C^{\rm CMB,N}$.  $C_g$ is the galaxy power
spectrum and $C_{g,N}$ is the associated measurement noise, $C_{g,N}=4\pi
f_{\rm sky}/N_g$ where $N_g$ is the total number of galaxies.  $f_{\rm sky}$
is the fractional sky coverage. 

We not only want the weighting to be unbiased ($\langle W\rangle_S=0$), but
also want it to be optimal such that the measurement error is minimized. It
turns out the unbiased optimal weighting is the solution to a  2nd integral
equation derived in the appendix \S \ref{sec:weighting}.  

The forecast for WMAP(PLANCK)-2MASS cross correlation measurement is shown in Fig. \ref{fig:SN}.  We take
the impact of galactic mask into account and adopt $f_{\rm  sky}=0.7$ to
evaluate the cosmic variance and shot  noise.  The unbiased optimal weighting
$W$ for this survey configuration is not trivial to derive. However, for the error
estimation presented here, it is safe to adopt the one corresponding to the case of full
sky coverage, for which we have the analytical solution
$W(\hat{n})=\cos\theta$.   Under such weighting we have $\langle
W^2\rangle_S=f_{Tg}=1/3$.    We find that
WMAP plus 2MASS is able to measure the dark flow, if it extends to $z\geq 0.1$
with an amplitude $10^3$ km/$s$, at $\sim 6\sigma$. Due to 
the resolution of WMAP, it will miss the peak of the correlation at $\ell\sim
10^3$. On the other hand, Planck, with better angular resolution, will be able
to well capture the peak correlation and is thus able to improve the measurement to
$\sim 14\sigma$. 

Since most 2MASS galaxies locate at $z<0.1$, 2MASS is inefficient to probe the dark flow at $z>0.1$. This explains the S/N plateau at
$z_{\rm DF}>0.1$ (Fig. \ref{fig:SN}). Deeper surveys such as SDSS are
more suitable for this purpose and can improve the overall S/N 
significantly. We expect that combining WMAP and existing 
galaxy surveys are already able to detect the dark flow at $z\la 0.3$ with
the claimed amplitude and depth at $\ga 6\sigma$ level. Planck will further
improve it to $\ga 14\sigma$, for $v_{\rm DF}=10^3$ km/$s$. Since the S/N
$\propto v_{\rm DF}$ in the cross correlation measurement, this implies that,
Planck plus existing galaxy 
surveys are able to detect dark flow with amplitude as low as $200$ km/$s$ at
$\ga 3\sigma$. 

\section{Summary}
We have pointed out the existence of  the dark flow induced small scale kinetic
SZ effect and estimated its amplitude.  As a potentially non-negligible 
component of CMB temperature fluctuations,  it impacts cosmology in at least
two ways. 
\bi
\item  It enables a useful independent check on
existing dark flow measurements 
\citep{Kashlinsky08,Kashlinsky09,Atrio-Barandela10,Kashlinsky10}. The
direction-weighted CMB-LSS cross correlation measurement proposed in this
paper should be able to detect a dark 
flow with an amplitude of $\ga 500$ km/$s$ at $\ga 3\sigma$ level for
WMAP+2MASS. Planck+2MASS can improve the detection threshold to $200$
km/$s$. Given this sensitivity, it will allow for a strong test of the
existence of the dark flow. 
\item A dark flow of the reported amplitude $\sim 10^3$ km/$s$ can
  significantly bias the ISW measurement through the CMB-LSS cross 
  correlation, if the survey area is sufficiently close to the direction of
  the dark flow or the opposite of it.   Hence it can significantly bias the
  dark energy constraint based on the ISW interpretation. Existing data shall
  be reinterpreted to avoid such bias. 
\ei

\section{Acknowledgment}
We thank Fernando Atrio-Barandela, Carlos Hern{\'a}ndez-Monteagudo and the
anonymous referee for useful discussions and suggestions. 
PJZ thanks the support of the one-hundred talents program of the Chinese
academy of science, the national science
foundation of China (grant No. 10821302 \& 10973027),   the CAS/SAFEA
International Partnership Program for  Creative Research Teams and the 973 program grant No. 2007CB815401.

\appendix
\section[]{Deriving the suppression factor}
\label{sec:cl}
We first subtract the dipole mode in  Eq. \ref{eqn:ksz} and rewrite the rest  as 
\be
\Theta_{\rm DF}=\cos\theta\times  y(\hat{n})\ ; \ y(\hat{n})\propto\int
\delta_e(1+z)^2d\tilde{x}\ .
\ee
The expectation value of
the angular correlation between two fixed directions $\hat{n}_1$ and $\hat{n}_2$ is 
\ba
\langle \Theta_{\rm DF}(\hat{n}_1)\Theta_{\rm DF}(\hat{n}_2)\rangle&=&
\cos\theta_1\cos\theta_2 \langle y(\hat{n_1})y(\hat{n}_2)\rangle\\
&=&\left[\cos^2\theta+O(\frac{\theta_{12}^2}{4})\right]\langle
y(\hat{n}_1)y(\hat{n}_2)\rangle\no\ .
\ea
Here, $\theta\equiv (\theta_1+\theta_2)/2$ and $\theta_{12}\equiv
\theta_1-\theta_2$. For small angular separation $|\Delta\theta|\ll 1$
($\hat{n}_1\cdot\hat{n}_2\equiv \cos\Delta \theta$), we have $|\theta_{12}|\leq
|\Delta \theta|\ll 1$. So the term $\theta_{12}^2/4\ll 1$ and can be safely
  neglected. Averaging over all pairs with the same $\Delta
  \theta$ in the survey volume and using the fact that $\langle
  y(\hat{n}_1)y(\hat{n}_2)\rangle$ should be isotropic, we have
\ba
\langle \Theta_{\rm DF}(\hat{n}_1)\Theta_{\rm DF}(\hat{n}_2)\rangle_S\simeq
\langle \cos^2\theta\rangle_S \langle y(\hat{n}_1)y(\hat{n}_2)\rangle\no\ .
\ea
Fourier transforming the above equation, we obtain Eq. \ref{eqn:cl} \&
Eq. \ref{eqn:fT} and recognize $f_T=\langle \cos^2\theta\rangle_S$ for the
case of $W=1$.  For a 
general $W$, $f_T=\langle \cos^2\theta W^2\rangle_S$ can be derived following
the same procedure. 

\section[]{The unbiased optimal weighting function}
\label{sec:weighting}
The optimal weighting $W(\hat{n})$ shall minimize $\langle W^2_S\rangle/f_{Tg}^2\propto
\langle W^2\rangle_S/\langle \cos\theta W\rangle^2_S$, under the constraint $\langle
W\rangle_S=0$.    By the Lagrange multiplier method, this corresponds to minimize
\be
\frac{\langle W^2\rangle_S}{\langle \cos\theta W\rangle_S^2}-\lambda \langle
W\rangle_S\ .
\ee
This requires 
\be
W=\frac{\langle W^2\rangle_S}{\langle \cos\theta W\rangle_S}\left[\cos\theta-\langle
\cos\theta\rangle_S\right)] \ .
\ee
For a general configuration of sky survey area, it is non-trivial to solve the
above second order integral equation. However, for the full sky coverage, since $\langle
\cos\theta \rangle_S=0$, one can easily find the solution to be $W=\cos\theta\equiv \hat{n}_{\rm
  DF}\cdot\hat{n}$.  This is what we adopt to estimate the expected WMAP+2MASS
cross correlation S/N. 

\end{document}